# A MEASUREMENT OF $\mathcal{B}(D_s^+ \to \phi l^+ \nu)/\mathcal{B}(D_s^+ \to \phi \pi^+)$

F. Butler,[1] X. Fu,[1] G. Kalbfleisch,[1] W.R. Ross,[1] P. Skubic,[1] J. Snow,[1] P.L. Wang,[1] M. Wood,[1] D.N. Brown,[2] J.Fast ,[2] R.L. McIlwain,[2] T. Miao,[2] D.H. Miller,[2] M. Modesitt,[2] D. Payne,[2] E.I. Shibata,[2] I.P.J. Shipsey,[2] P.N. Wang,[2] M. Battle,[3] J. Ernst,[3] Y. Kwon,[3] S. Roberts,[3] E.H. Thorndike,[3] C.H. Wang,[3] J. Dominick,[4] M. Lambrecht,[4] S. Sanghera,[4] V. Shelkov,[4] T. Skwarnicki,[4] R. Stroynowski,[4] I. Volobouev,[4] G. Wei,[4] P. Zadorozhny,[4] M. Artuso,[5] M. Goldberg,[5] D. He,[5] N. Horwitz,[5] R. Kennett,[5] R. Mountain,[5] G.C. Moneti,[5] F. Muheim,[5] Y. Mukhin,[5] S. Playfer,[5] Y. Rozen,[5] S. Stone,[5] M. Thulasidas,[5] G. Vasseur,[5] G. Zhu,[5] J. Bartelt,[6] S.E. Csorna,[6] Z. Egyed,[6] V. Jain,[6] K. Kinoshita,[7] K.W. Edwards,[8] M. Ogg,[8] D.I. Britton,[9] E.R.F. Hyatt,[9] D.B. MacFarlane,[9] P.M. Patel,[9] D.S. Akerib,[10] B. Barish,[10] M. Chadha,[10] S. Chan,[10] D.F. Cowen,[10] G. Eigen,[10] J.S. Miller,[10] C. O'Grady,[10] J. Urheim,[10] A.J. Weinstein,[10] D. Acosta,[11] M. Athanas,[11] G. Masek,[11] H.P. Paar,[11] J. Gronberg,[12] R. Kutschke,[12] S. Menary,[12] R.J. Morrison,[12] S. Nakanishi,[12] H.N. Nelson,[12] T.K. Nelson,[12] C. Qiao,[12] J.D. Richman,[12] A. Ryd,[12] H. Tajima,[12] D. Sperka,[12] M.S. Witherell,[12] M. Procario,[13] R. Balest,[14] K. Cho,[14] M. Daoudi,[14] W.T. Ford,[14] D.R. Johnson,[14] K. Lingel,[14] M. Lohner,[14] P. Rankin,[14] J.G. Smith,[14] J.P. Alexander,[15] C. Bebek,[15] K. Berkelman,[15] K. Bloom,[15] T.E. Browder,[15*] D.G. Cassel,[15] H.A. Cho,[15] D.M. Coffman,[15] P.S. Drell,[15] R. Ehrlich,[15] P. Gaiderev,[15] M. Garcia-Sciveres,[15] B. Geiser,[15] B. Gittelman,[15] S.W. Gray,[15] D.L. Hartill,[15] B.K. Heltsley,[15] C.D. Jones,[15] S.L. Jones,[15] J. Kandaswamy,[15] N. Katayama,[15] P.C. Kim,[15] D.L. Kreinick,[15] G.S. Ludwig,[15] J. Masui,[15] J. Mevissen,[15] N.B. Mistry,[15] C.R. Ng,[15] E. Nordberg,[15] J.R. Patterson,[15] D. Peterson,[15] D. Riley,[15] S. Salman,[15] M. Sapper,[15] F. Würthwein,[15] P. Avery,[16] A. Freyberger,[16] J. Rodriguez,[16] R. Stephens,[16] S. Yang,[16] J. Yelton,[16] D. Cinabro,[17] S. Henderson,[17] T. Liu,[17] M. Saulnier,[17] R. Wilson,[17] H. Yamamoto,[17] T. Bergfeld,[18] B.I. Eisenstein,[18] G. Gollin,[18] B. Ong,[18] M. Palmer,[18] M. Selen,[18] J. J. Thaler,[18] A.J. Sadoff,[19] R. Ammar,[20] S. Ball,[20] P. Baringer,[20] A. Bean,[20] D. Besson,[20] D. Coppage,[20] N. Copty,[20] R. Davis,[20] N. Hancock,[20] M. Kelly,[20] N. Kwak,[20] H. Lam,[20] Y. Kubota,[21] M. Lattery,[21] J.K. Nelson,[21] S. Patton,[21] D. Perticone,[21] R. Poling,[21] V. Savinov,[21] S. Schrenk,[21] R. Wang,[21] M.S. Alam,[22] I.J. Kim,[22] B. Nemati,[22] J.J. O'Neill,[22] H. Severini,[22] C.R. Sun,[22] M.M. Zoeller,[22] G. Crawford,[23] C. M. Daubenmier,[23] R. Fulton,[23] D. Fujino,[23] K.K. Gan,[23] K. Honscheid,[23] H. Kagan,[23] R. Kass,[23] J. Lee,[23] R. Malchow,[23] Y. Skovpen,[23†] M. Sung,[23] and C. White[23]

(CLEO Collaboration)

[1] University of Oklahoma, Norman, Oklahoma 73019
[2] Purdue University, West Lafayette, Indiana 47907
[3] University of Rochester, Rochester, New York 14627
[4] Southern Methodist University, Dallas, Texas 75275
[5] Syracuse University, Syracuse, New York 13244
[6] Vanderbilt University, Nashville, Tennessee 37235
[7] Virginia Polytechnic Institute and State University, Blacksburg, Virginia, 24061
[8] Carleton University, Ottawa, Ontario K1S 5B6 and the Institute of Particle Physics, Canada
[9] McGill University, Montréal, Québec H3A 2T8 and the Institute of Particle Physics, Canada
[10] California Institute of Technology, Pasadena, California 91125
[11] University of California, San Diego, La Jolla, California 92093
[12] University of California, Santa Barbara, California 93106
[13] Carnegie-Mellon University, Pittsburgh, Pennsylvania 15213
[14] University of Colorado, Boulder, Colorado 80309-0390
[15] Cornell University, Ithaca, New York 14853
[16] University of Florida, Gainesville, Florida 32611
[17] Harvard University, Cambridge, Massachusetts 02138
[18] University of Illinois, Champaign-Urbana, Illinois, 61801
[19] Ithaca College, Ithaca, New York 14850
[20] University of Kansas, Lawrence, Kansas 66045
[21] University of Minnesota, Minneapolis, Minnesota 55455
[22] State University of New York at Albany, Albany, New York 12222
[23] Ohio State University, Columbus, Ohio, 43210

## Abstract

Using the CLEO II detector at CESR, we have measured the ratio of branching fractions $\mathcal{B}(D_s^+ \to \phi e^+ \nu)/\mathcal{B}(D_s^+ \to \phi \pi^+) = 0.54 \pm 0.05 \pm 0.04$. We use this measurement to obtain a model dependent estimate of $\mathcal{B}(D_s^+ \to \phi \pi^+)$.

---

*Permanent address: University of Hawaii at Manoa

†Permanent address: INP, Novosibirsk, Russia

Most measurements of the $D_s^+$ meson branching fractions are normalized to the clean $D_s^+ \to \phi\pi^+$ channel. [1] However, the absolute $D_s^+ \to \phi\pi^+$ branching fraction is not well known, and this limits the precision of these measurements. Here we present a measurement of $R_s = \mathcal{B}(D_s^+ \to \phi l^+\nu)/\mathcal{B}(D_s^+ \to \phi\pi^+)$, which can be used to extract the $D_s^+ \to \phi\pi^+$ branching fraction by using the known values for the $D \to \bar{K}^* l\nu$ branching fractions, the $D$ and $D_s^+$ meson lifetimes and theoretical predictions for the ratio of widths: $\Gamma(D_s^+ \to \phi l^+\nu)/\Gamma(D \to \bar{K}^* l^+\nu)$.

The data consist of an integrated luminosity of 1.71 $\text{fb}^{-1}$ of $e^+e^-$ collisions recorded with the CLEO II detector at the Cornell Electron Storage Ring (CESR). A detailed description of the CLEO II detector can be found in reference [2]. The data sample contains over two million $e^+e^- \to c\bar{c}$ events taken at center-of-mass energies on the $\Upsilon(4S)$ resonance and in the nearby continuum ($\sqrt{s} \sim 10.6$ GeV).

Due to the undetected neutrino, we cannot fully reconstruct $D_s^+ \to \phi l^+\nu$ decays. However, there are very few processes which produce both a $\phi$ meson and a lepton contained in the same jet. Consequently, this correlation can be used to extract a clean $D_s^+ \to \phi l^+\nu$ signal. The backgrounds due to misidentified leptons and from random $\phi$–lepton combinations can be reliably estimated, and the possible contamination from other decay modes is shown to be negligible.

We identify $\phi$ candidates by using the decay mode $\phi \to K^+K^-$. In order to suppress combinatoric background, $\phi$ candidates are required to have momenta above 1.1 GeV/c. In addition, the kaon candidates must have ionization energy loss and time-of-flight consistent with that expected for a kaon with the measured momentum.

The search for leptons is restricted to the kinematic regions in which the lepton identification efficiencies and hadron misidentification rates are well understood. Hence, electron and muon candidates are required to be in the fiducial regions $|\cos\theta| < 0.91$ and $|\cos\theta| < 0.81$, respectively, where $\theta$ is the polar angle of the track with respect to the beam-axis. In addition, electron candidates must have momenta above 0.9 GeV/c and muon candidates above 1.4 GeV/c. The only exception is for muons in the region $|\cos\theta| > 0.61$ which are required to have momenta above 1.9 GeV/c. Electrons are identified by comparing their ionization energy loss, time-of-flight, and energy deposited in the electromagnetic calorimeter with that expected for an electron with the measured momentum. Electrons from photon conversions and Dalitz decays of $\pi^0$'s are rejected by pairing electron candidates with all other oppositely charged tracks in the event and rejecting those which have both small separation and parallel trajectories at their point of closest approach. Muons are identified by matching charged tracks to hits in the muon detectors which lie outside the electromagnetic calorimeter. In order to be identified as a muon, a track must penetrate at least 5 interaction lengths of steel. For leptons in the momentum ranges and fiducial regions considered, the identification efficiencies are approximately 92% for electrons and 90% for muons.

To reduce further the combinatoric background, we require that the $\phi l^+$ momentum be greater than 2.4 GeV/c. In order to be consistent with having originated from a $D_s^+$ decay, the $\phi l^+$ candidates must have an invariant mass less than 1.9 GeV/c$^2$. In order to suppress the combinatoric background from $\Upsilon(4S)$ events which tend to be more spherical, we require that the ratio of Fox-Wolfram moments [3], $R_2 = H_2/H_0$, is greater than 0.30. This eliminates 77% of the $\Upsilon(4S)$ background whilst retaining 92% of the signal.

The efficiencies for reconstructing $D_s^+ \to \phi l^+ \nu$ decays are obtained from a Monte Carlo simulation which takes the predictions of the ISGW model [4] as input. These events are then passed through a full simulation of the CLEO II detector and the same event reconstruction and analysis chain as the real data. Because of the small $q^2$ value associated with the decay $\phi \to K^+K^-$, the kaons tend to overlap in the drift-chamber. This makes it difficult to simulate accurately the ionization energy loss measurement. In order to avoid this problem, the momentum dependent efficiencies for identifying $\phi$ mesons are obtained from the data by comparing the inclusive yield of all $\phi$'s before and after particle identification. These efficiencies are then combined with the predicted $\phi$ momentum spectrum from $D_s^+ \to \phi l^+ \nu$ decays to give the total $\phi$ identification efficiency. Following the above selection criteria, and after correcting for the effects of final-state radiation from the leptons [5], the efficiencies for identifying $D_s^+ \to \phi e^+ \nu$ decay is 10.5% and for $D_s^+ \to \phi \mu^+ \nu$ the efficiency is 2.9%.

Figs 1(a) and 1(b) show the invariant mass distributions of all $K^+K^-$ combinations which are accompanied by an electron or muon respectively and which pass the above selection criteria. We fit these distributions with a signal and background function. The signal function is a Gaussian function convoluted with a Breit-Wigner function. The background function is a phase-space background function [6] which accounts for random $K^+K^-$ combinations. The width of the Breit-Wigner function is fixed to the natural width of the $\phi$ state [8], and the mean and sigma of the Gaussian function are fixed to the values extracted from a fit to all $\phi$ candidates with momenta above 1.1 GeV/c. Only the overall normalization of the signal function is allowed to vary in the fits. The fits yield $359 \pm 22$ $D_s^+ \to \phi e^+ \nu$ and $123 \pm 15$ $D_s^+ \to \phi \mu^+ \nu$ candidates. There are two main sources of background: $\phi$'s accompanied by fake leptons [7], and random $\phi l^+$ combinations.

The background due to fake leptons is estimated by first using the real data to measure the momentum dependent probabilities that a hadron will be misidentified as a lepton. These probabilities are typically 0.3% for electrons and 1.2% for muons. These results are then used to randomly label tracks (which do not pass the lepton identification criteria described above) as leptons in the data. With this procedure, we extract the number of $\phi l^+$ combinations due to misidentified hadrons. For electrons this estimate is $46 \pm 14$ events, while for muons it is $27 \pm 8$. The quoted errors include the contributions from the uncertainties in the misidentification probabilities. In order to check these estimates, we examine the invariant mass distribution of all $\phi l^+$ candidates. A peak at the $D_s^+$ mass is seen which is due to $D_s^+ \to \phi \pi^+$ decays in which the pion is misidentified as either an electron or muon. The technique of randomly labeled hadronic tracks as leptons yields $7.4 \pm 2.2$ $\phi\pi$ events where the $\pi$ track is misidentified as a lepton. This is in good agreement with the $4.5 \pm 3.6$ events found in the lepton sample.

For the range of lepton momenta considered, random $\phi l^+$ combinations come from two sources: from $e^+e^- \to c\bar{c}$ events in which a $\phi$ is produced in the fragmentation process and is combined with a lepton from the semileptonic decay of the charmed hadron in the same jet, and from $\Upsilon(4S)$ decays in which a $\phi$ is produced in the decay chain of one of the $B$ mesons and is combined with a lepton from the semileptonic decay of the other $B$ meson. A $\phi$ and lepton which originate from the decay of the same $B$ meson do not contribute since they tend to be emitted back-to-back and thus have too large an invariant mass. The background from random $\phi l^+$ combinations is estimated using the Monte Carlo simulation. However, this is complicated by the fact that the $\phi$ production rate from both fragmentation and $B$ meson decays is not well known. For this reason, an attempt is made to scale the Monte Carlo prediction to account for the $\phi$ production rate observed in the data.

In the continuum, not only the rate of $\phi$ production, but also the correlation between the $\phi$ and the charmed hadron direction is important. The agreement between the data and Monte Carlo is investigated by considering how often a $\phi$ is produced in the same hemisphere as a fully reconstructed $D$ meson. Both $D^0$ and $D^{*+}$ mesons are considered, and are reconstructed using the following decay chains: $D^0 \to K^-\pi^+$ and $D^{*+} \to D^0\pi^+$; $D^0 \to K^-\pi^+$. The reconstructed $D$ mesons are required to have momenta above 2.5 GeV/c in order to account approximately for the range of $D$ momenta which are expected to contribute leptons in the momentum range of interest. For this particular study, the $\phi$ momentum criterion is relaxed to 0.8 GeV/c in order to provide sufficient statistics. In doing this we have assumed that the $\phi$ momentum distribution is well reproduced by the Monte Carlo, and that it is the rate of $\phi$ production in the fragmentation process which contributes the greatest uncertainty. Both the number of $D$ mesons and the number $\phi$'s are obtained by fitting their invariant mass distributions. False combinations due to the $D$ meson combinatoric backgrounds are accounted for by subtracting the number of $\phi$'s found when using the $D$ mesons invariant mass sidebands. In the real data $0.17 \pm 0.11$ $\phi$'s are found for every 1000 reconstructed $D$ mesons. This is to be compared with $0.16 \pm 0.02$ for the $e^+e^- \to c\bar{c}$ Monte Carlo. The ratio of these two numbers is $1.0 \pm 0.7$, so no correction is applied in this case. The simulation predicts a background of $12 \pm 8$ and $1.8 \pm 1.2$ events for electrons and muons respectively, where the errors include the uncertainty in the above ratio.

The background from random $\phi l^+$ combinations in $\Upsilon(4S)$ decays is estimated in a similar manner. In this case the directions of the $\phi$ and lepton are uncorrelated. For this reason it is sufficient to compare the number of $\phi$'s with momentum above 1.1 GeV/c in the continuum subtracted $\Upsilon(4S)$ data with that observed in the $\Upsilon(4S)$ $B\bar{B}$ Monte Carlo. In the real data $5.0 \pm 0.5$ $\phi$'s are found per 1000 $B\bar{B}$ events to be compared with $5.3 \pm 0.1$ in the Monte Carlo. This gives a correction factor of $0.95 \pm 0.08$. After applying this correction, the predicted background is $19 \pm 2$ events for electrons and $9 \pm 1$ events for muons.

Figs 2(a) and 2(b) show the number of $\phi$'s which fall in each $\phi l^+$ invariant mass bin for electrons and muons respectively; where the number of $\phi$'s has been extracted from fits to the $K^+K^-$ invariant mass distributions. The combined background estimates are also shown, as well as the simulated predictions for the signal shapes which have been normalized to the number of candidates extracted from the fits to the $K^+K^-$ invariant mass spectra. It can be seen that the predicted signal shapes are in good agreement with the data. The background estimates can be checked by comparing the predicted number of candidates which fall outside of the signal region with the number actually observed. For electrons we predict $8 \pm 1$ events in the region $2.0 < M_{\phi l^+} < 3.5$ GeV/c$^2$ and observe $12 \pm 7$, and for muons we predict $7 \pm 1$ and observe $8 \pm 5$. Both predictions are in good agreement with the data.

We have also estimated the possible contamination from the decays $D^+ \to \phi \bar{K}^0 l^+ \nu$, $D_s^+ \to \phi \eta l^+ \nu$ and $D_s^+ \to \phi \pi \pi l^+ \nu$. The decay $D_s^+ \to \phi \pi^0 l^+ \nu$ is forbidden from conservation of isospin. The Feynman diagrams for the first two processes are shown in Figs. 3(a) and 3(b), respectively. For the first decay, we first estimate an upper limit on the number of $D^+ \to \phi \bar{K}^0 l^+ \nu$ decays in the data by making the conservative assumption: $\mathcal{B}(D^+ \to \phi \bar{K}^0 l^+ \nu)$ $/\mathcal{B}(D^+ \to \bar{K}^0 l^+ \nu) \simeq \alpha \cdot \mathcal{B}(D^0 \to (K^* \pi)^- \mu^+ \nu)/\mathcal{B}(D^0 \to K^- \mu^+ \nu)$, where $\alpha$ is a suppression factor because in the first numerator it is an $s\bar{s}$ pair which must be popped from the vacuum as opposed to a light-quark pair [9]. This assumption is motivated by the similarity of the decay diagrams. The latter ratio has been measured by the E653 collaboration [10], $\mathcal{B}(D^0 \to (K^*\pi)^-\mu^+\nu)/\mathcal{B}(D^0 \to K^-\mu^+\nu) < 0.04$ at the 90% confidence level. Taking the number of $D^+ \to \bar{K}^0 l^+\nu$ events in our data sample to be 60,000 [11], assuming $\alpha = 1/3$, and including the simulated acceptance for $D^+ \to \phi \bar{K}^0 l^+\nu$ decays, we estimate less than one background event from this source. The contribution from $D_s^+ \to \phi\eta l^+\nu$ decays is estimated in a similar manner. Here we make use of the same E653 result and assume: $\mathcal{B}(D_s^+ \to \phi\eta l^+\nu)/\mathcal{B}(D_s^+ \to \phi l^+\nu) \simeq \alpha \cdot \beta \cdot \mathcal{B}(D^0 \to (K^*\pi)^-\mu^+\nu)/\mathcal{B}(D^0 \to K^-\mu^+\nu)$, where $\beta$ accounts for OZI suppression [12] of the first numerator. Making the assumption $\beta = 1/10$, and including the simulated acceptance for $D_s^+ \to \phi\eta l^+\nu$ decays, we again estimate much less than one background event from this source. The contribution from $D_s^+ \to \phi\pi\pi l^+\nu$ decays should also be very small for similar reasons. Therefore, it is assumed that the background from these decay modes is negligible.

After subtracting all backgrounds, we find $282 \pm 22$ $D_s^+ \to \phi e^+\nu$ and $85 \pm 15$ $D_s^+ \to \phi\mu^+\nu$ candidates which fall in the $D_s^+$ signal region $M_{\phi l^+} < 1.9$ GeV/c$^2$. After correcting for the detection efficiencies in each channel and for the $\phi \to K^+K^-$ branching fraction [8], the efficiency corrected yields are $5460 \pm 430$ for electrons and $6000 \pm 1000$ for muons. A breakdown of the yields in each channel is given in Table I. To combine these two numbers, we take a weighted average, after first correcting for the fact that the muon rate is predicted to be 5% lower than that for electrons because of the reduced phase-space [13]. Therefore, our result is given in terms of the effective yield in the electron channel which is $5580 \pm 400$ events.

In order to limit the systematic effects which stem from the selection criteria, the number of $D_s^+ \to \phi\pi^+$ decays is measured in a similar manner. Again, the $\phi$ candidates are required to have momenta above 1.1 GeV/c . To account approximately for the fact that no neutrino is produced in this decay, we require the $\phi\pi^+$ momentum to be greater than 2.7 GeV/c [14]. We then require the $K^+K^-\pi^+$ invariant mass to be within $\pm 25$ MeV/c$^2$ of the known $D_s^+$ mass [8]. The efficiency for detecting $D_s^+ \to \phi\pi^+$ decays following these selection criteria is

17.4%.

The result of the fit to the $K^+K^-$ invariant mass distribution is shown in Fig. 4. We find $1049 \pm 35$ candidates. Also shown is the result of the fit to $K^+K^-$ combinations from the $D_s^+$ mass sidebands, which is used to estimate the contribution from random $\phi\pi^+$ combinations. We find $163 \pm 18$ candidates due to these random combinations. After subtracting this background, the efficiency corrected yield is $10,370 \pm 460$ events.

Finally, since we have already corrected for efficiencies, the ratio of branching fractions is,

$$R_s = \frac{\mathcal{B}(D_s^+ \to \phi e^+ \nu)}{\mathcal{B}(D_s^+ \to \phi\pi^+)} = \frac{5580 \pm 400}{10,370 \pm 460} = 0.54 \pm 0.05 \pm 0.04, \tag{1}$$

where the first error is statistical, and the second is an estimate of possible systematic effects. This systematic error includes: the uncertainty in the number of fake leptons (6.3%), the uncertainty in the level of continuum charm background (2.7%), the uncertainty in the level of $B\bar{B}$ background (0.9%), the uncertainty in the lepton identification efficiency (2.5%), the uncertainty in the $\phi$ identification efficiency (1.0%) and that due to the limited number of Monte Carlo events which were used for the efficiency estimates (2.7%). We have also considered our sensitivity to the $D_s^+$ production mechanism by using the predicted efficiencies for $D_s^+$ mesons produced in $D_s^{*+}$ decays. However, since both the $D_s^+ \to \phi l^+ \nu$ and the $D_s^+ \to \phi\pi^+$ efficiencies are affected in the same manner, the effect is small (0.5%). For this analysis the ISGW model was used to generate semileptonic decays in the Monte Carlo simulation. The uncertainty associated with this choice of model was investigated by adopting the $(V - A)$ prediction for the lepton momentum spectra. From the resulting change in the predicted $D_s^+ \to \phi l^+ \nu$ acceptance, we assign a systematic error of 1.8%. Various background shapes have also been used to extract the number of $\phi$ mesons. In all cases $R_s$ changes by less than 1.1%, which is taken to be the systematic error. After adding these estimates in quadrature, the total systematic error is 7.6%. Table II compares this result with those of previous measurements [15–17].

Having measured $R_s$, we extract the $D_s^+ \to \phi\pi^+$ branching fraction by using the theo-retical value for $\mathcal{F}_s = \Gamma(D_s^+ \to \phi l^+ \nu)/\Gamma(D \to \bar{K}^* l^+ \nu)$. We can write

$$\mathcal{B}(D_s^+ \to \phi e^+ \nu) = \mathcal{F}_s \times \mathcal{B}(D^0 \to K^{*-} e^+ \nu) \times \frac{\tau_{D_s^+}}{\tau_{D^0}}$$
$$= \mathcal{F}_s \times \frac{\mathcal{B}(D^0 \to K^{*-} e^+ \nu)}{\mathcal{B}(D^0 \to K^- \pi^+)} \times \mathcal{B}(D^0 \to K^- \pi^+) \times \frac{\tau_{D_s^+}}{\tau_{D^0}}. \quad (2)$$

where $\tau_{D_s^+}$ and $\tau_{D^0}$ are the $D_s^+$ and $D^0$ lifetimes, respectively. $\mathcal{F}_s$ is predicted by various quark models [18,19]. Here we choose to adopt the result of the modified ISGW model $\mathcal{F}_s = 1.00$, since to date this is the only model which can account for the measured value for $\mathcal{B}(D \to \bar{K}^* l^+ \nu)/\mathcal{B}(D \to K l^+ \nu)$ [19,20]. Past experiments have used $\mathcal{F}_s = 0.9$ for this ratio. We use CLEO II measurements for all quantities except the $D^0$ and $D_s^+$ lifetimes. In this way some of the systematic errors cancel and the problems associated with averaging the results of many different experiments are avoided. We found $\mathcal{B}(D^0 \to K^{*-} e^+ \nu)/\mathcal{B}(D^0 \to K^- \pi^+) = 0.61 \pm 0.07$ [11], and when this is combined with our measurement, $\mathcal{B}(D^0 \to K^- \pi^+) = (3.91 \pm 0.19)\%$ [21], and with the E687 measurements, $\tau_{D^0} = (4.13 \pm 0.05) \times 10^{-13}$ s [22] and $\tau_{D_s^+} = (4.75 \pm 0.21) \times 10^{-13}$ s [23], we obtain $\mathcal{B}(D_s^+ \to \phi e^+ \nu) = (2.74 \pm 0.36)\%$. Using our measurement of $R_s$, we obtain: $\mathcal{B}(D_s^+ \to \phi \pi^+) = (5.1 \pm 0.4 \pm 0.4 \pm 0.7)\%$, where the first error is from the statistical error on $R_s$, the second from the systematic error on $R_s$, and the third from the uncertainty in the $D_s^+ \to \phi e^+ \nu$ branching fraction. This measurement is greater than but consistent with previous estimates and upper limits on $\mathcal{B}(D_s^+ \to \phi \pi^+)$ [8,24].

In conclusion, we have measured $\mathcal{B}(D_s^+ \to \phi e^+ \nu)/\mathcal{B}(D_s^+ \to \phi \pi^+) = 0.54 \pm 0.05 \pm 0.04$. By using the theoretical prediction $\mathcal{F}_s = 1.00$, we find $\mathcal{B}(D_s^+ \to \phi \pi^+)$= $(5.1 \pm 0.4 \pm 0.4 \pm 0.7)\%$.

We gratefully acknowledge the effort of the CESR staff in providing us with excellent luminosity and running conditions. J.P.A. and P.S.D. thank the PYI program of the NSF, I.P.J.S. thanks the YI program of the NSF, G.E. thanks the Heisenberg Foundation, K.K.G., I.P.J.S., and T.S. thank the TNRLC, K.K.G., H.N.N., J.D.R., T.S. and H.Y. thank the OJI program of DOE and P.R. thanks the A.P. Sloan Foundation for support. This work was supported by the National Science Foundation and the U.S. Dept. of Energy.

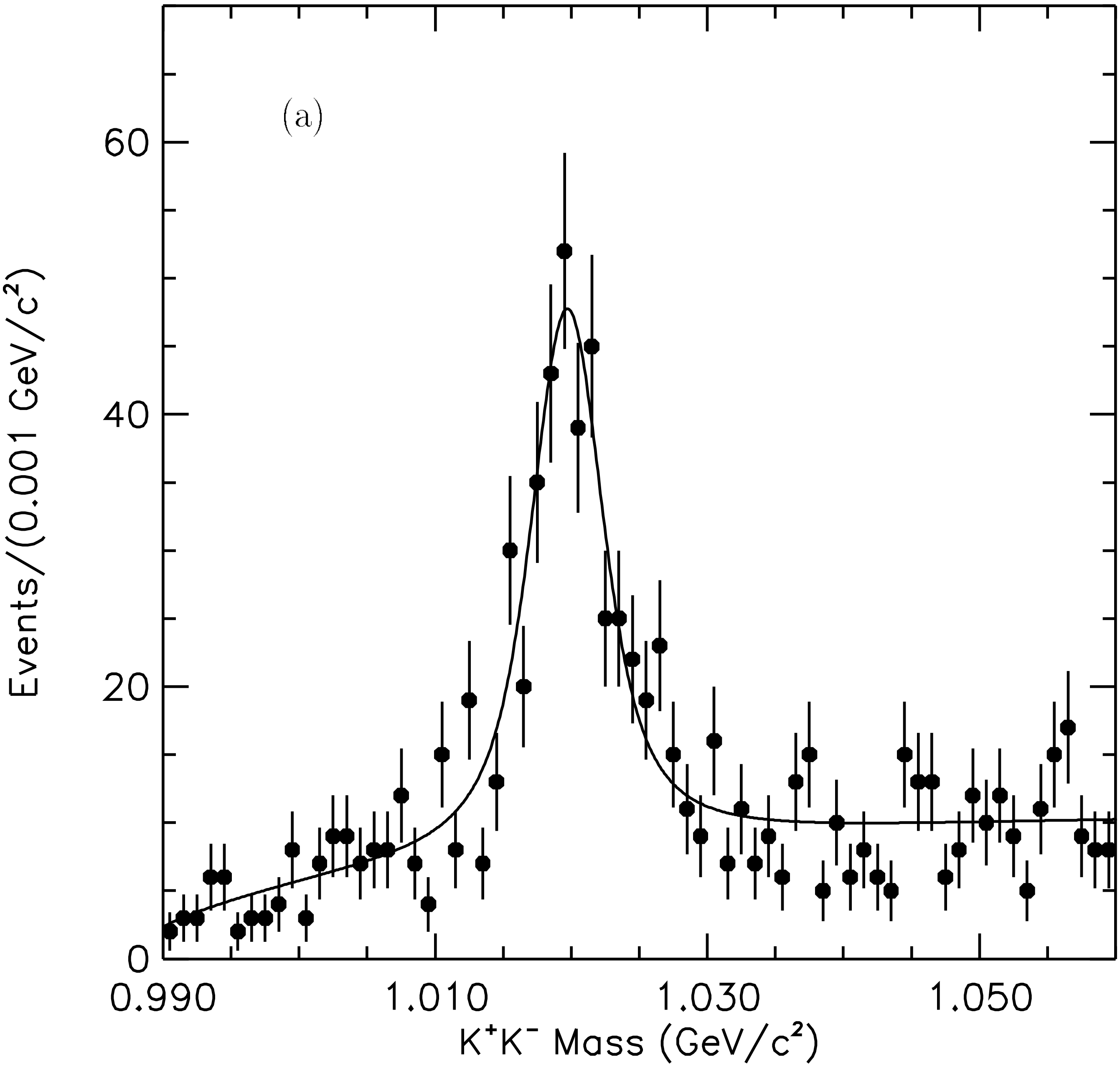

FIG. 1. Fits to the $K^+K^-$ invariant mass distributions for (a) $K^+K^-e^+$ and (b) $K^+K^-\mu^+$ combinations which lie in the $D_s^+$ signal region $M_{K^+K^-l^+} < 1.9$ GeV/c$^2$.

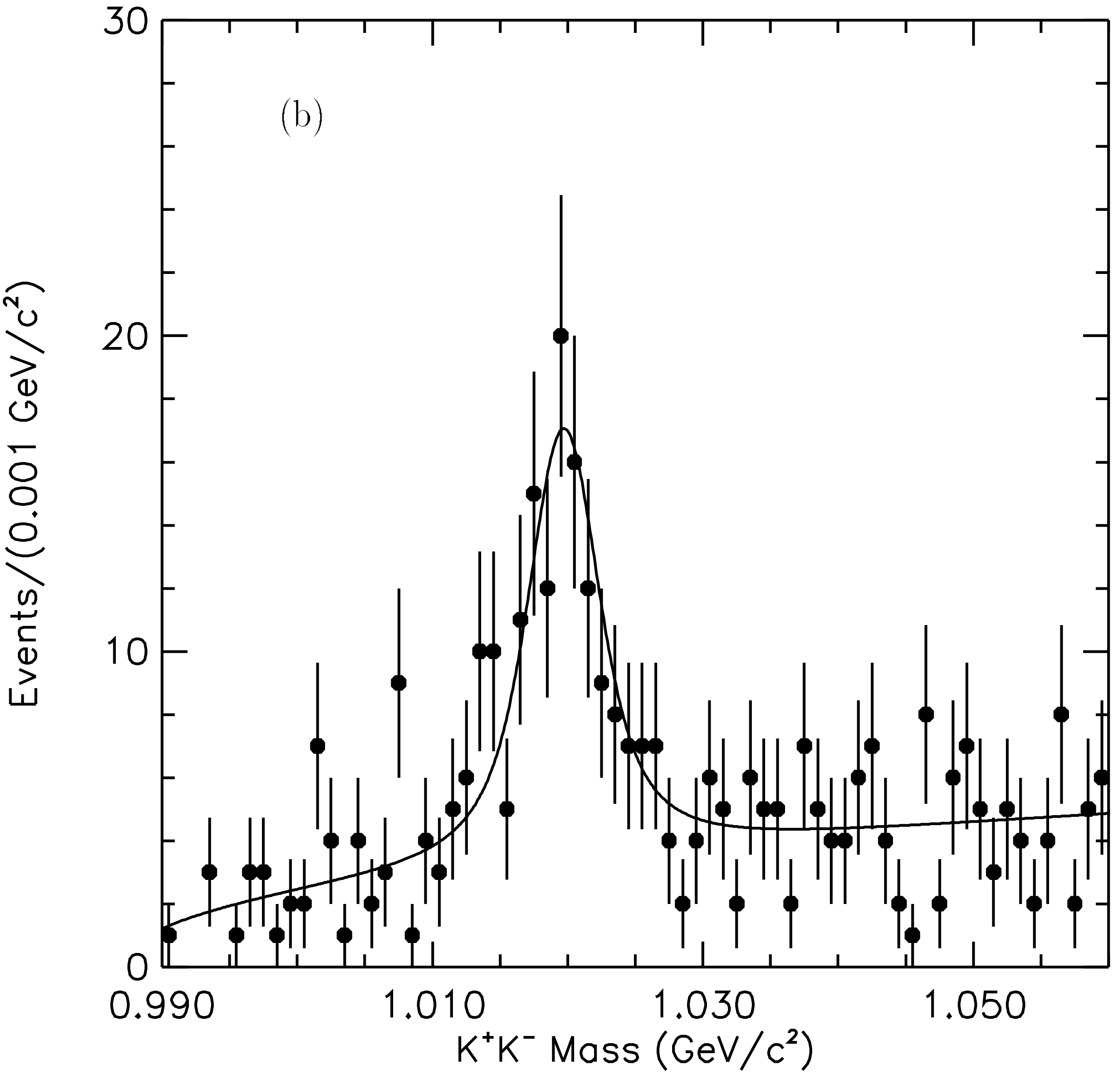

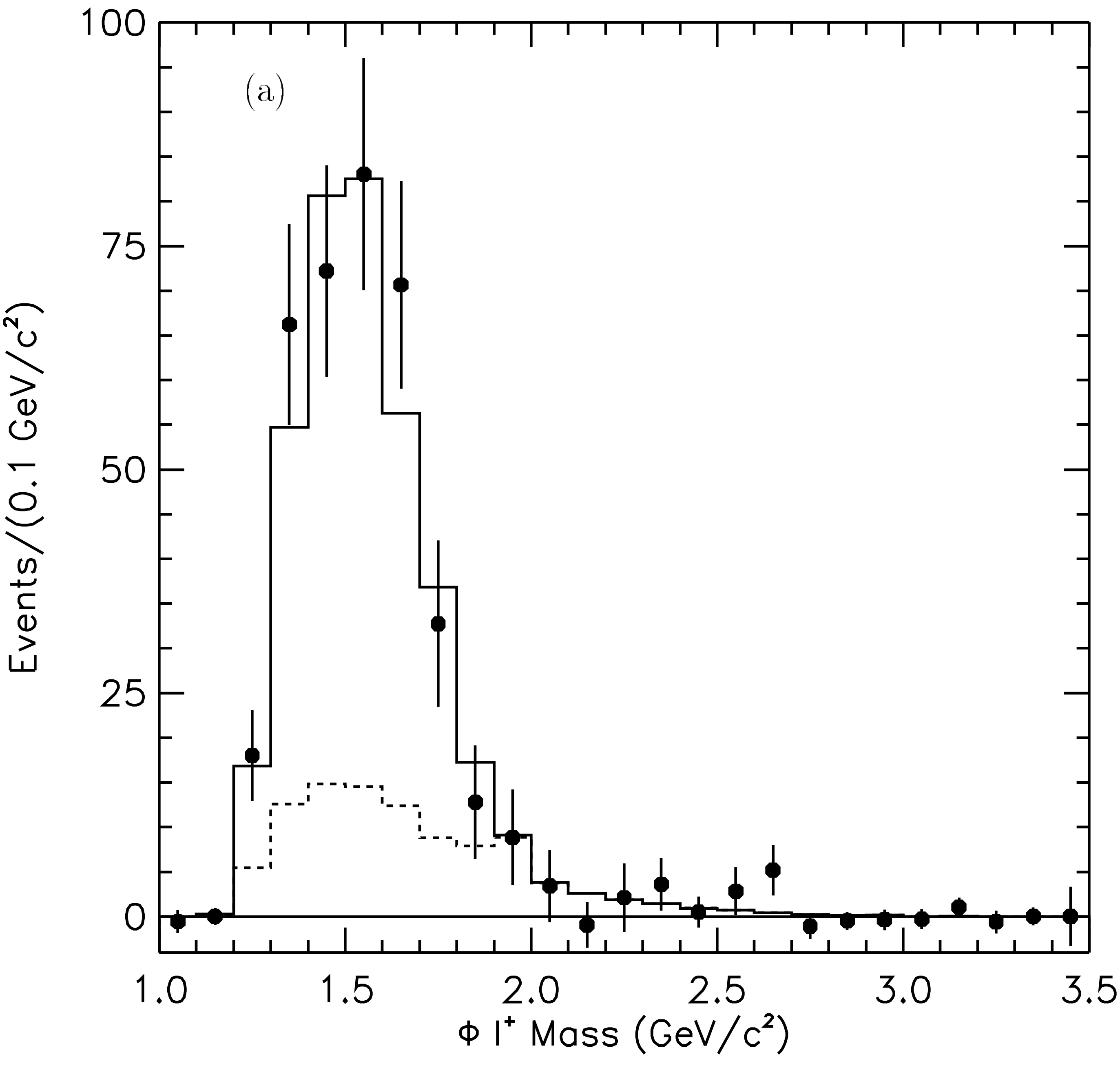

FIG. 2. Invariant mass of (a) $\phi e^+$ and (b) $\phi\mu^+$ combinations. The data points are obtained by fitting the $K^+K^-$ invariant mass distributions for each $\phi l^+$ invariant mass bin. The solid histograms show the sums of the predicted backgrounds and the simulated signal shapes. The dashed histograms show the background contributions.

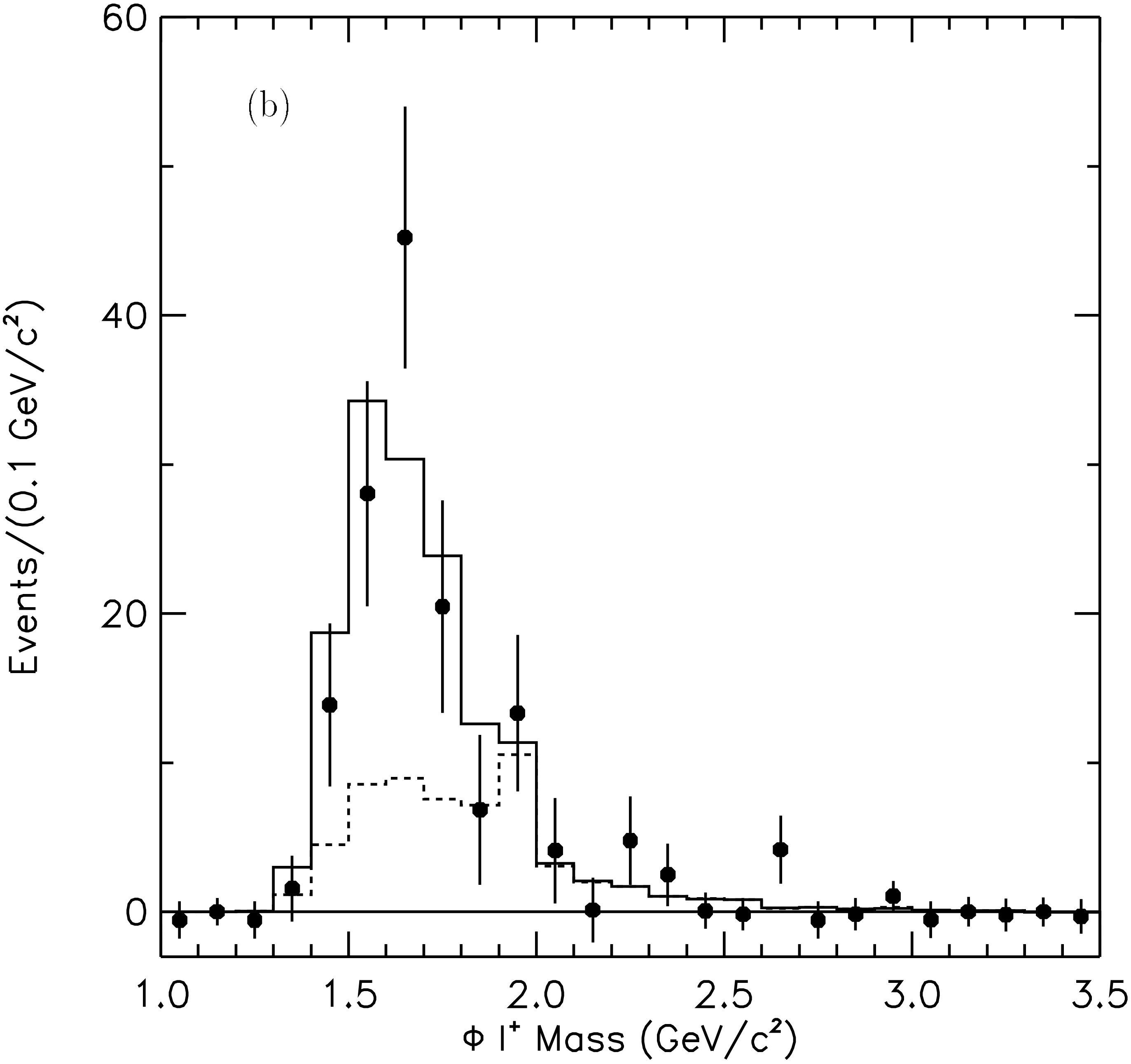

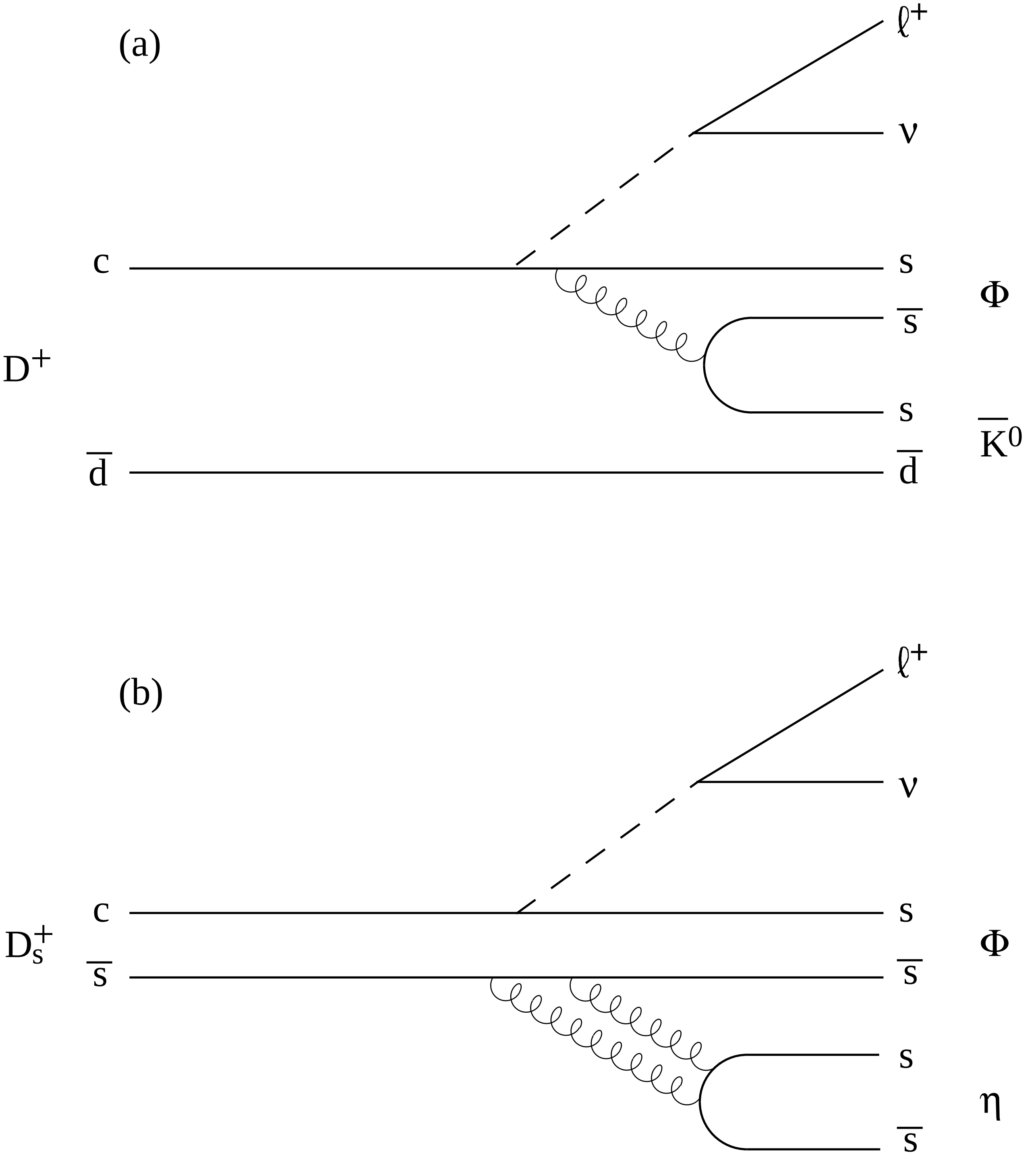

FIG. 3. Feynman diagrams for the possible background modes. (a) $D^+ \to \phi \bar{K^0} l^+ \nu$, in which an $s\bar{s}$ pair must be "popped" from the vacuum, and (b) $D_s^+ \to \phi \eta l^+ \nu$, which is OZI suppressed.

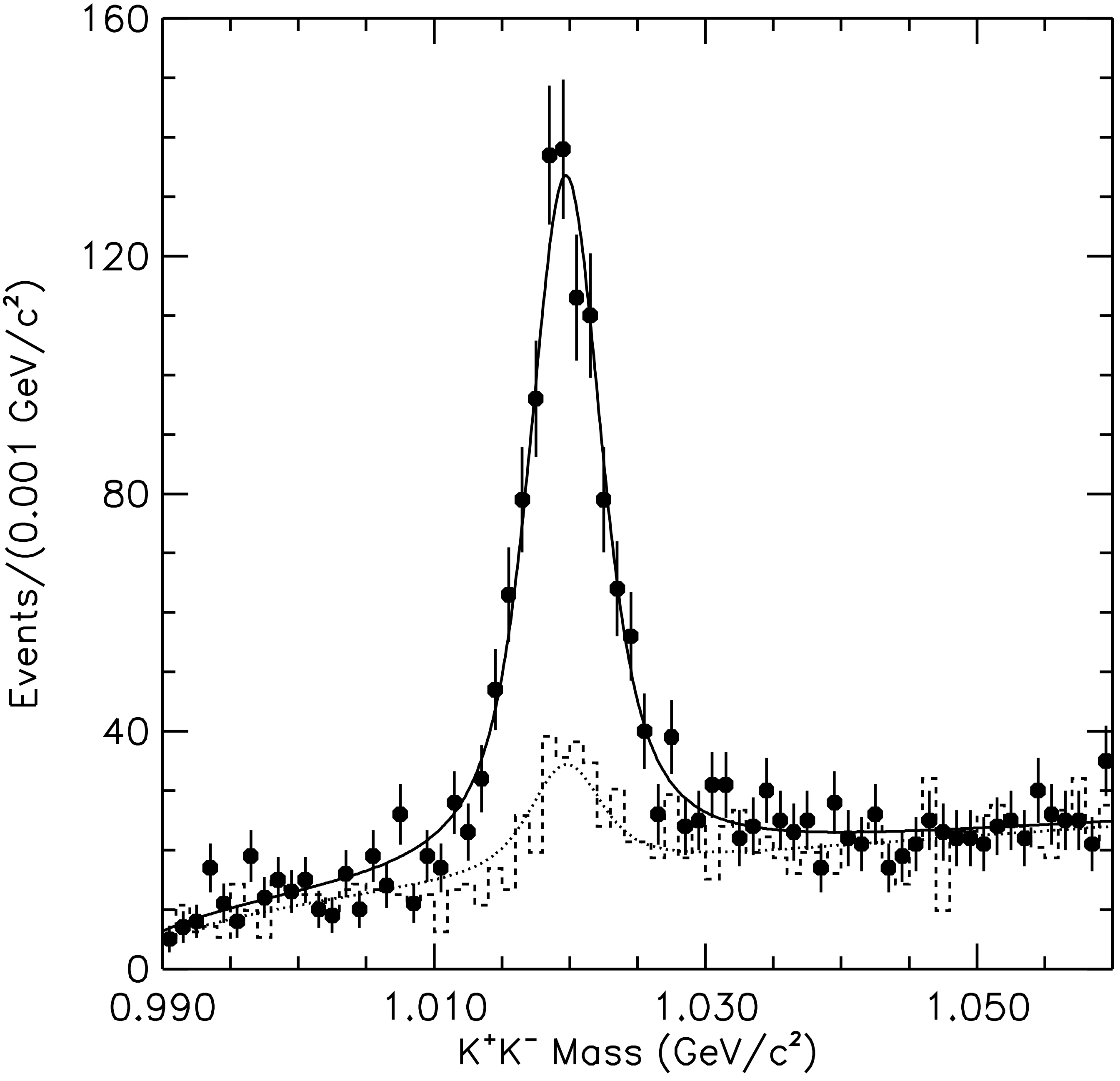

FIG. 4. Fit to the $K^+K^-$ invariant mass distribution for $K^+K^-\pi^+$ combinations which lie within $\pm 25$ MeV/c$^2$ of the $D_s^+$ mass. The dashed histogram shows the contribution from the $D_s^+$ mass sidebands.

# TABLES

TABLE I. Summary of $D_s^+ \to \phi l^+ \nu$ yields. The errors quoted in this table are statistical only.

| Decay mode | $D_s^+ \to \phi e^+ \nu$ | $D_s^+ \to \phi \mu^+ \nu$ |
|---|---|---|
| Total candidates | $359 \pm 22$ | $123 \pm 15$ |
| Fake lepton background | $46 \pm 0.3$ | $27 \pm 0.8$ |
| Continuum $c\bar{c}$ background | $12 \pm 0.4$ | $1.8 \pm 0.1$ |
| $B\bar{B}$ background | $19 \pm 0.8$ | $9 \pm 0.5$ |
| Background subtracted | $282 \pm 22$ | $85 \pm 15$ |
| Efficiency, $\epsilon \cdot \mathcal{B}$ (%) | 5.16 | 1.42 |
| Efficiency corrected yield | $5460 \pm 430$ | $6000 \pm 1000$ |

TABLE II. Comparison of this result with those of previous experiments. We have increased the E687 result by 5% since only muons were used in their analysis.

| Experiment | Events | $R_S$ |
|---|---|---|
| CLEO 1.5 [15] | 54 | $0.49 \pm 0.10^{+0.10}_{-0.14}$ |
| ARGUS [16] | 104 | $0.57 \pm 0.15 \pm 0.15$ |
| E687 [17] | 97 | $0.61 \pm 0.18 \pm 0.07$ |
| This result | 367 | $0.54 \pm 0.05 \pm 0.04$ |